\documentclass[onecolumn]{mn2e}
\newif\ifAMStwofonts

\def\pmb#1{\mbox{\boldmath$#1$}}
\def\gtsim {>\kern-1.2em\lower1.1ex\hbox{$\sim$}}
\def\ltsim {<\kern-1.2em\lower1.1ex\hbox{$\sim$}}
\def\gtsim {>\kern-1.2em\lower1.1ex\hbox{$\sim$}}
\def\ltsim {<\kern-1.2em\lower1.1ex\hbox{$\sim$}}
\def\ref{\hangindent=1pc \hangafter=1 \noindent}

\def\be{\begin{equation}}
\def\ee{\end{equation}}

\usepackage{epsfig}
\begin{document}



\title{Light curves from rapidly rotating neutron stars}
\author[K. Numata \& U. Lee]{Kazutoshi Numata$^1$\thanks{E-mail: numata@astr.tohoku.ac.jp; Current address: Hitachi Software Engineering Co.,Ltd., Shinagawa, Tokyo 140-0002, Japan} \& Umin Lee$^1$\thanks{E-mail: lee@astr.tohoku.ac.jp}
\\$^1$Astronomical Institute, Tohoku University, Sendai, Miyagi 980-8578, Japan}

\date{Typeset \today ; Received / Accepted}
\maketitle


\begin{abstract}
We calculate light curves produced by a hot spot of a rapidly rotating neutron star, assuming that
the spot is perturbed by a core $r$-mode, which is destabilized by emitting gravitational
waves.
To calculate light curves, we take account of relativistic effects such as the Doppler
boost due to the rapid rotation and light bending assuming the Schwarzschild metric around the neutron star.
We assume that the core $r$-modes
penetrate to the surface fluid ocean to have sufficiently large amplitudes to disturb the spot.
For a $l'=m$ core $r$-mode, the oscillation frequency $\omega\approx2m\Omega/[l'(l'+1)]$
defined in the co-rotating frame of the star will be detected by a distant observer, 
where $l'$ and $m$ are respectively the spherical harmonic degree and the azimuthal wave number of the mode, and $\Omega$ is the spin frequency of the star.
In a linear theory of oscillation, using a parameter $A$
we parametrize the mode amplitudes such that ${\rm max}\left(|\xi_\theta|,|\xi_\phi|\right)/R=A$ at the surface, 
where $\xi_\theta$ and $\xi_\phi$ are the $\theta$ and $\phi$ components of the displacement vector of the mode
and $R$ is the radius of the star.
For the $l'=m=2$ $r$-mode with $\omega=2\Omega/3$,
we find that the fractional Fourier amplitudes at $\omega=2\Omega/3$ in light curves
depend on the angular distance $\theta_s$ of the spot centre measured from the rotation axis and 
become comparable to or even larger than $A\sim0.001$ for small values of $\theta_s$.
\end{abstract}


\section{Introduction}

Accretion powered millisecond X-ray pulsars in low mass X-ray binaries (LMXBs)
show small amplitude, almost sinusoidal X-ray time variations,
the dominant periods of which are thought to correspond to the spin periods of the neutron stars
(e.g., Lamb \& Boutloukos 2007).
Lamb et al (2009) argued that the millisecond X-ray variations
are produced by an X-ray emitting hot spot located at a magnetic pole of the rotating
neutron star, and that so long as the center of the hot spot is only slightly off the rotation axis
the X-ray variations produced will have small amplitudes and become almost sinusoidal.
They also suggested that if the hot spot is located close to the rotation axis,
a slight drift of the hot spot away from the rotation axis leads to appreciable changes in the amplitudes and phases of the X-ray variations.
Lamb et al (2009) pointed out that a temporal
change in mass accretion rates and hence the radius of the magnetosphere, for example, can cause 
such a drift of the hot spot.

It is now well known that neutron stars can support various kinds of oscillation modes
(e.g., McDermott et al 1988).
There have been, however, no observational evidences that definitely indicate global oscillations of the stars, probably except for quasi-periodic oscillations observed in the tail of 
giant X-ray flares from Soft Gamma-ray Repeaters (SGRs) (Duncan 1998; Israel et al 2005; Strohmayer \& Watts 2005, 2006; Watts \& Strohmayer 2006), which are
believed to have a global magnetic field as strong as $B\gtsim10^{14}$G at the surface (e.g., Woods \& Thompson 2006).
The QPOs observed in SGRs may be caused by damping oscillations excited when a sudden restructuring of
a global magnetic field in the neutron star takes place.
For any global oscillations of a neutron star to become observable, mechanisms are needed that excite the oscillations
to have amplitudes large enough to produce observable variations in the radiation flux.
The $\epsilon$ mechanism can be an example of such excitation mechanisms
for low frequency $g$-modes and $r$-modes in the surface fluid layers of mass accreting
neutron stars (e.g., Strohmayer \& Lee 1996; Narayan \& Cooper 2007).
In fact, the $r$-modes propagating in the surface fluid layers of mass accreting neutron stars
have been proposed for the burst oscillations observed during type I X-ray bursts in low mass X-ray binaries (Strohmayer et al 1996, 1997; Heyl 2004, 2005; Lee 2004; Lee \& Strohmayer 2005).
It is also argued that retrograde oscillation modes can be destabilized by emitting gravitational waves if they satisfy the frequency condition given by $0<\omega/\Omega<m$ (Friedman \& Schutz 1978), and that $r$-modes, which are retrograde modes, are to become unstable to the gravitational radiation reaction (Andersson 1998), where $\Omega$ is the spin frequency of the star, $\omega$ is the oscillation frequency observed in the corotating frame, and $m$ is the azimuthal wave number of the mode.
$R$-modes in the fluid core of rapidly rotating neutron stars have been a subject of intensive studies
(e.g., Lindblom, Owen, \& Morsink 1998; Owen et al. 1998; Lockitch \& Friedman 1999; Yoshida \& Lee 2000ab, 2001)
and are now regarded as a possible candidate for oscillation modes of a neutron star that are excited to
produce observable effects.

If a hot spot on a rapidly rotating neutron star can produce clean light curves without
any strong harmonics of the spin frequency,
it may be possible to use the light curves as a probe into oscillation modes that are excited to
periodically disturb the spot so that the periodicities of the modes manifest themselves in the light curves.
In this paper, we calculate light curves by a hot spot taking account of the disturbances by $r$-modes
in the fluid core, which are assumed to be excited by emitting gravitational waves.
Although the amplitudes of the $r$-modes are confined into the core, 
in the presence of a global magnetic field the amplitudes may penetrate to the surface
fluid ocean (e.g., Lee 2010).
In general, the radial component of the displacement vector of the $r$-modes at the surface have amplitudes
much smaller that the horizontal and toroidal components, and hence the temperature
variations at the surface produced by the $r$-modes would be too small to have observable amplitudes.
If the horizontal component of the $r$-modes is large enough to deform the hot spot appreciably, on the other hand,
the periodic deformation of the spot may manifest itself in the light curves produced by the hot spot.
\S 2 is for method of solution, and numerical results are given in \S 3, and \S 4
is for conclusion.

\section{Method of calculation}

We consider a rapidly rotating neutron star having a hot spot on the surface and 
assume that $r$-modes in the fluid core are destabilized by emitting gravitational waves (Friedman \& Morsink 1998; Andersson 1998).
For a distant observer, light curves produced by the hot spot will then be observed to have
periodic flux variations with the dominant period equal to the spin period of the star.
If the core $r$-modes excited by emitting gravitational waves can penetrate 
the solid crust to have sufficient amplitudes at the surface to give periodic disturbances 
to the spot, it is expected that the periodicities due to the core $r$-modes will be contained
in the light curves from the hot spot.
It is the aim of this section to present a method of calculation of light curves produced
by a surface hot spot which is disturbed by a core $r$-mode.

It is convenient to use two Cartesian coordinate systems $\left(x,y,z\right)$ and $\left(x',y',z'\right)$
and to assume the distant observer is in the $x$-$z$ and $x'$-$z'$ planes,
where the $z$-axis is the spin axis of the star and $z'$-axis is pointing to the distant
observer and $y=y'$.
We denote by $(r,\theta,\phi)$ and $(r',\theta',\phi')$ spherical polar coordinates associated with
the Cartesian coordinates, where the $z$-axis ($z'$-axis) corresponds to the axis defined by $\theta=0$
($\theta'=0$).
We denote by $i$ the inclination angle between the $z$-axis and $z'$-axis.
If we assume Schwarzschild metric to calculate 
photon trajectories around the spinning neutron star,
for a photon emitted from a point $\left(R,\theta',\phi'\right)$ on the surface of the star 
and reaching the distant observer, the angle $\theta'$ is
given by (see Pechenick, Ftaclas \& Cohen 1983)
\be
\theta'=\int_R^\infty{dr'\over r'^2}\left[{1\over b^2}-{1\over r'^2}\left(1-{r_g\over r'}\right)\right]^{-1/2},
\ee
where $b$ is the impact parameter given by
\be
b={R\over\sqrt{1-r_g/R}}\sin\delta
\ee
where $\delta$ is the angle between the surface normal vector $\pmb{n}$ and the direction vector $\pmb{l}$, measured by a non-rotating observer at the stellar surface, of the photon that reaches the distant observer, and
$r_g=2GM/c^2$ is the Schwarzschild radius, and $R$ and $M$ are the radius and mass of the star, and 
$G$ and $c$ are the gravitational constant and
the light velocity, respectively.
Note that the three vectors $\pmb{n}$, $\pmb{l}$, and $\pmb{k}'$ are coplanar, where
$\pmb{k}'$ is the unit vector along the $z'$-axis.
The observed differential flux $dF_E$ may be given by (e.g., Poutanen \& Gierli\'nski 2003)
\be
dF_E=I_EdO={\sqrt{1-r_g/R}\over D^2}\eta^3 \hat I_{\hat E}\left(\hat \delta\right){d\cos\delta\over d\cos\theta'}\cos\hat\delta d\hat S,
\ee
where $I_E$ and $dO$ are the intensity of radiation at energy $E$ and the solid angle
seen by a distant observer, $D$ is the distance to the observer from the centre of the star, $\hat I_{\hat E}\left(\hat \delta\right)$ is the intensity of radiation at energy $\hat E$ into the direction angle $\hat\delta$ measured from the surface normal, and $d\hat S$ is the area of a surface element on the surface
and the hatted quantities indicate those defined in the frame corotating with the star, and we have used
$\cos\delta dS=\cos\hat\delta d\hat S$ (e.g., Ghisellini 1999; Lind \& Blandford 1985).
Here, $\eta$ is the Doppler factor given by
\be
\eta={1\over\gamma\left(1-\beta\cos\zeta\right)},
\ee
where 
\be 
\gamma={1\over \sqrt{1-\beta^2}}, \quad
\beta={R\Omega/c\over\sqrt{1-r_g/R}}\sin\theta,
\quad {\rm and} \quad
\cos\zeta=-{\sin\delta\over\sin\theta'}\sin i\sin\phi.
\ee
Integrating the flux $dF_E$ by photon energy $E$ 
measured by a distant observer, we obtain
\be
dF
={\left(1-r_g/R\right)\over D^2}\eta^5\cos\delta {d\cos\delta\over d\cos\theta'}
\hat I\left(\hat\delta\right) d\hat S,
\ee
where we have used 
\be
E=\eta\sqrt{1-r_g/R}\hat E, \quad \eta\cos\delta=\cos\hat\delta, \quad {\rm and} 
\quad \hat I\left(\hat\delta\right)=\int_0^\infty\hat I_{\hat E}\left(\hat \delta\right)d\hat E.
\ee
Assuming $\hat I(\hat\delta)=\hat I_0$ is homogneous black body radiation independent of $\hat\delta$ and integrating over the surface in the corotating frame, 
we have
\be
F={\left(1-r_g/R\right)\hat I_0}{R^2\over D^2}\int_{\hat S} \eta^5 {\sin\delta\over \sin\theta'} { d\sin\delta\over d\theta'}
 \sin\hat\theta d\hat\theta d\hat\phi,
\ee
where $(\hat r,\hat\theta,\hat\phi)$ are spherical polar coordinates in the corotating frame of the star, and we assume $\theta=\hat\theta$ and $\phi=\hat\phi+\Omega t$ with
$\Omega$ being the spin frequency of the star
observed by a distant observer, and $t$ is the coordinate time at infinity.
Hereafter, we let $(\hat r,\hat\theta,\hat\phi)$ and $(\hat x,\hat y,\hat z)$ denote coordinates
defined in the corotating frame of the star.

To take account of the effects of periodic disturbances due to core $r$-modes on the spot,
we first calculate small amplitude oscillations of rotating and magnetized neutron stars in Newtonian dynamics, disregarding general relativistic effects on the oscillations (Lee 2010).
We introduce a dipole magnetic field given by $\pmb{B}=\mu_m\nabla(\cos\theta/r^2)$ whose 
magnetic axis is assumed to align with the spin axis of the star,
where $\mu_m$ is the magnetic dipole moment.
Since the magnetic pressure in the deep interior is much smaller than the gas pressure
for neutron stars with a magnetic field whose strength at the surface is 
comparable to or less than $\sim10^{12}$G, we treat the fluid core as being non-magnetic (e.g., Lee 2007, 2010).
Since we assume the spin axis is the magnetic axis,
the temporal and azimuthal angular dependence of oscillations can be represented by a single factor 
${\rm e}^{{\rm i}(m\hat\phi+\omega t)}$,
where $m$ is the azimuthal wavenumber around the rotation axis and $\omega\equiv\sigma+m\Omega$ 
is the oscillation frequency in the corotating frame of the star with
$\sigma$ being the oscillation frequency in an inertial frame.
Since the angular dependence of the oscillations in a rotating and magnetized star cannot be represented by
a single spherical harmonic function, we expand the perturbed quantities in terms of 
spherical harmonic functions $Y_l^m\left(\theta,\phi\right)$ with different $l$s for a given $m$, 
considering that the axis of rotation coincides with that of the magnetic field.
The displacement vector $\pmb{\xi}$ is then represented by a finite series expansion of length $j_{\rm max}$ as
\be
{\pmb{\xi}\over \hat r}=\sum_{j=1}^{j_{\rm max}}\left[S_{l_j}(\hat r)Y^m_{l_j}(\hat\theta,\hat\phi)\hat{\pmb{e}}_r
+H_{l_j}(\hat r)\nabla_{\rm H} Y^m_{l_j}(\hat\theta,\hat\phi)
+T_{l^\prime_j}(\hat r)~\hat{\pmb{e}}_r\times\nabla_{\rm H} Y^m_{l^\prime_j}(\hat\theta,\hat\phi)\right]{\rm e}^{{\rm i}\omega t},
\ee
where $\hat{\pmb{e}}_r$, $\hat{\pmb{e}}_\theta$, and $\hat{\pmb{e}}_\phi$ are the orthonormal vectors in the $\hat r$, $\hat \theta$,
and $\hat\phi$ directions, respectively, and 
\be
\nabla_{\rm H}=\hat{\pmb{e}}_\theta{\partial\over\partial\hat \theta}+\hat{\pmb{e}}_\phi{1\over\sin\hat\theta}{\partial\over\partial\hat\phi},
\ee
and $l_j=|m|+2(j-1)$ and $l^\prime_j=l_j+1$ for even modes, and 
$l_j=|m|+2j-1$ and $l^\prime_j=l_j-1$ for odd modes, respectively, and $j=1,~2,~3,~\cdots, ~j_{\rm max}$.
Substituting such expansions into the linearized basic equations,
we obtain a finite set of coupled linear ordinary differential equations for the expansion coefficients
$S_{l_j}$, $H_{l_j}$, and $T_{l^\prime_j}$ (e.g., Lee 2007, 2010).
When the angular dependence of the displacement vector at the surface is
represented by the functions $\Xi_j\left(\hat\theta\right)$ defined by
\be
\Xi_r\left(\hat\theta\right)e^{{\rm i}m\hat\phi}=\sum _{j=1}^{j_{\rm max}}S_{l_j}\left(R\right)Y_{l_j}^m\left(\hat\theta,\hat\phi\right)
\ee
\be
\Xi_\theta\left(\hat\theta\right)e^{{\rm i}m\hat\phi}=\hat{\pmb{e}}_\theta\cdot\sum_{j=1}^{j_{\rm max}}\left[
H_{l_j}(R)\nabla_{\rm H} Y^m_{l_j}(\hat\theta,\hat\phi)
+T_{l^\prime_j}(R)~\hat{\pmb{e}}_r\times\nabla_{\rm H} Y^m_{l^\prime_j}(\hat\theta,\hat\phi)\right],
\ee
\be
\Xi_\phi\left(\hat\theta\right)e^{{\rm i}m\hat\phi}=-{\rm i}\hat{\pmb{e}}_\phi\cdot\sum_{j=1}^{j_{\rm max}}\left[
H_{l_j}(R)\nabla_{\rm H} Y^m_{l_j}(\hat\theta,\hat\phi)
+T_{l^\prime_j}(R)~\hat{\pmb{e}}_r\times\nabla_{\rm H} Y^m_{l^\prime_j}(\hat\theta,\hat\phi)\right],
\ee
we can rewrite the displacement vector at the surface as
\be
\pmb{\xi}/R=\left[\Xi_r\left(\hat\theta\right)\hat{\pmb{e}}_r+\Xi_\theta\left(\hat\theta\right)\hat{\pmb{e}}_\theta
+{\rm i}\Xi_\phi\left(\hat\theta\right)\hat{\pmb{e}}_\phi\right]\exp{\rm i}\left(m\hat\phi+\omega t\right),
\ee
the real part of which is given by
\be
{\rm Re}\left(\pmb{\xi}\right)/R=\left[\Xi_r\left(\hat\theta\right)\hat{\pmb{e}}_r+\Xi_\theta\left(\hat\theta\right)\hat{\pmb{e}}_\theta\right]\cos\left(m\hat\phi+\omega t\right)-\Xi_\phi\left(\hat\theta\right)\hat{\pmb{e}}_\phi\sin\left(m\hat\phi+\omega t\right).
\ee
Note that $m\hat\phi+\omega t=m\phi+\sigma t$.
In Figure 1, the functions $\Xi_r$, $\Xi_\theta$, and $\Xi_\phi$ of the $l'=m=2$ core $r$-mode
calculated for a $0.5M_\odot$ neutron star model composed of a fluid core, a solid crust, and a
fluid ocean are plotted, from the left to right panels, for $\bar\Omega\equiv\Omega/\sqrt{GM/R^3}=0.1$, 0.2, and 0.3,
where we have assumed $B_0=10^{10}$G with $B_0$ being the strength of
a dipole magnetic field at the surface (see Lee 2010), and in each panel
the amplitudes of the functions are normalized by ${\rm max}(|\Xi_r|,|\Xi_\theta|,|\Xi_\phi|)$.
As shown by the figure, since the amplitudes of the function $\Xi_r$ at the surface are much smaller than those of
$\Xi_\theta$ and $\Xi_\phi$ for the $r$-modes, we neglect the
term $\Xi_r$ in the following.
In a linear theory of stellar oscillations,
the amplitudes of the oscillations are indeterminate, and we have to 
treat the amplitudes as a parameter.
In this paper, using a parameter $A$ we normalize the oscillation amplitudes such that ${\rm max}\left(|\Xi_\theta(\hat\theta)|,|\Xi_\phi(\hat\theta)|\right)=A$ in $0\le\hat\theta\le\pi$.

\begin{figure}
\resizebox{0.33\columnwidth}{!}{
\includegraphics{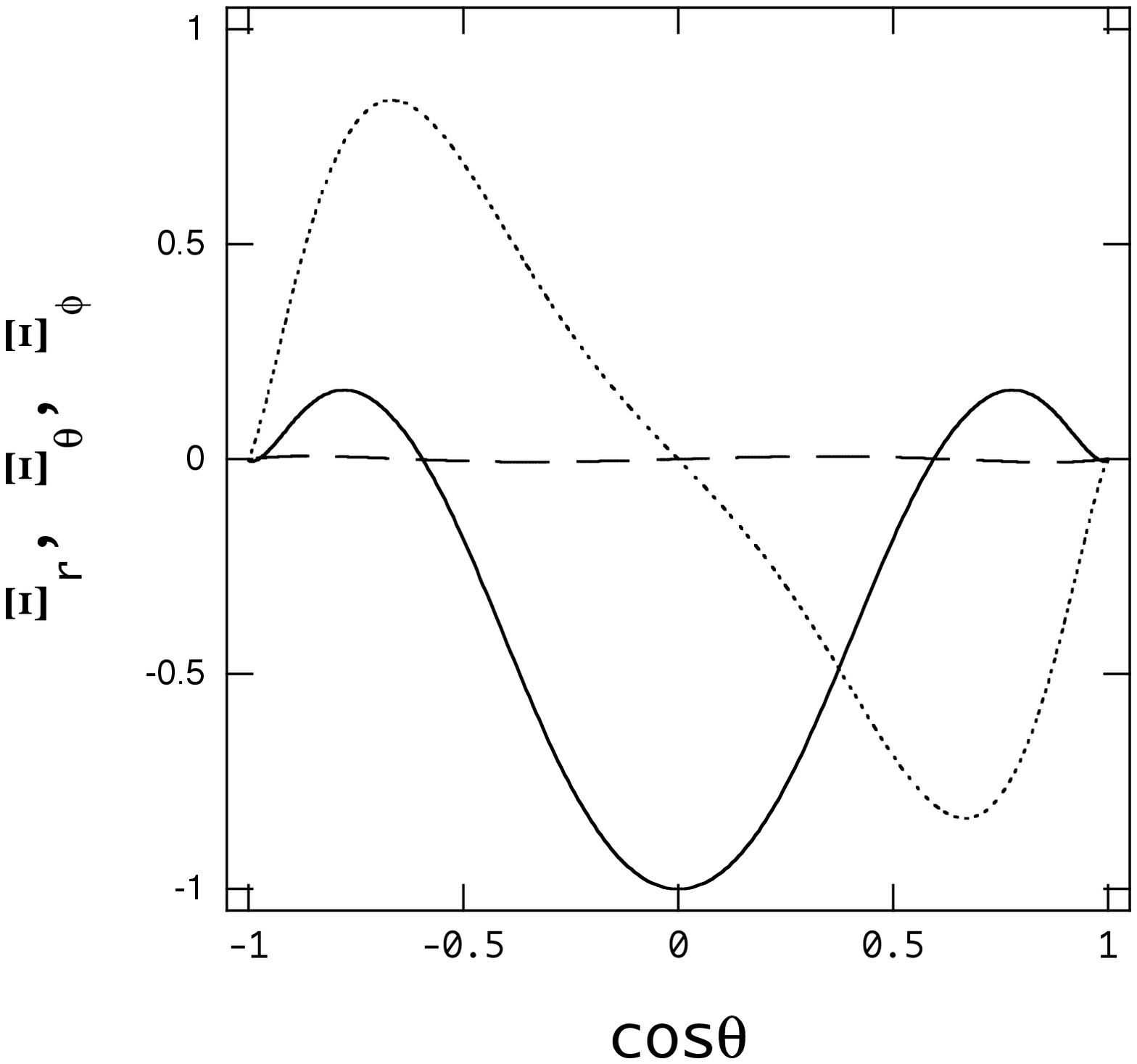}}
\resizebox{0.33\columnwidth}{!}{
\includegraphics{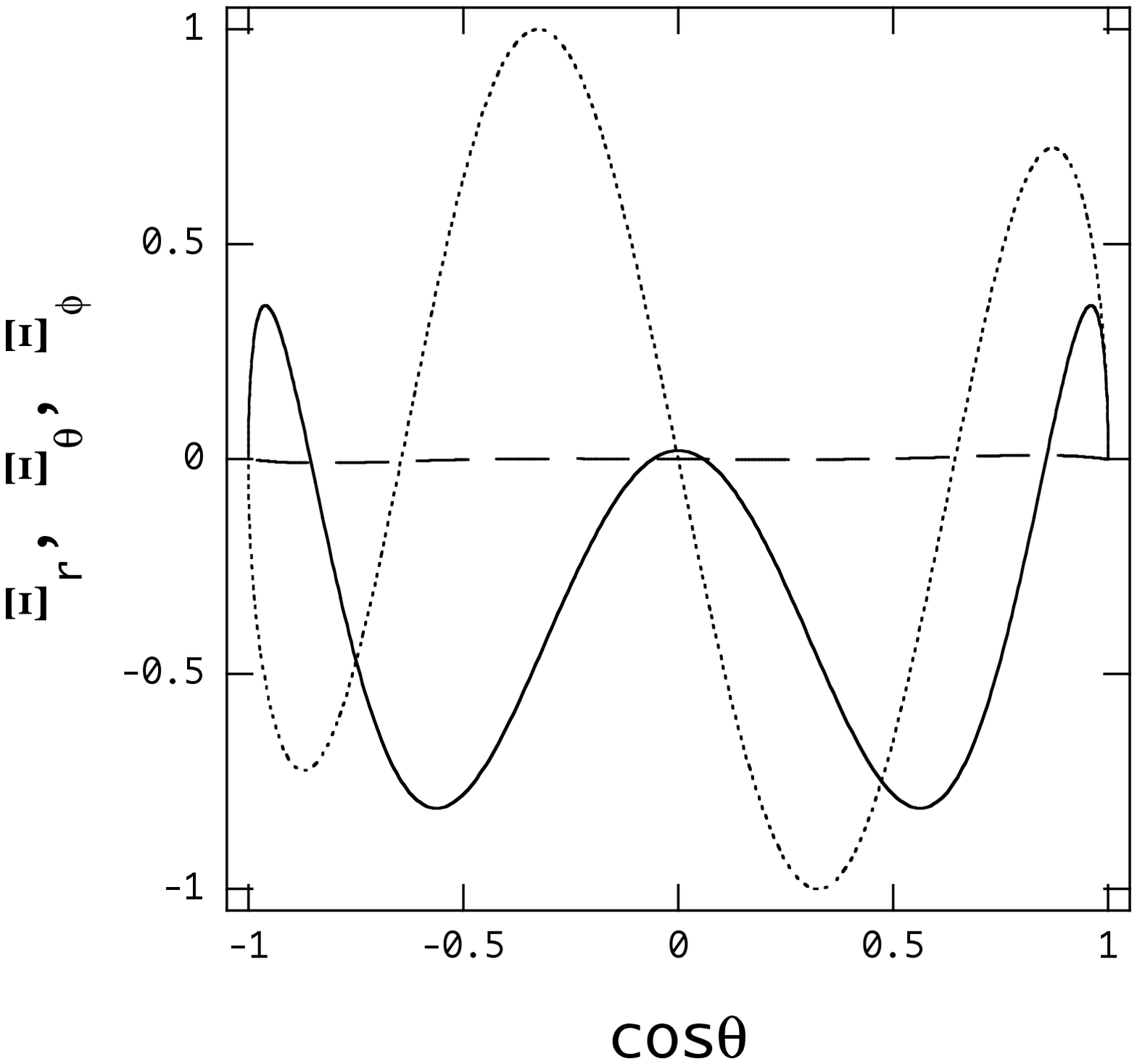}}
\resizebox{0.33\columnwidth}{!}{
\includegraphics{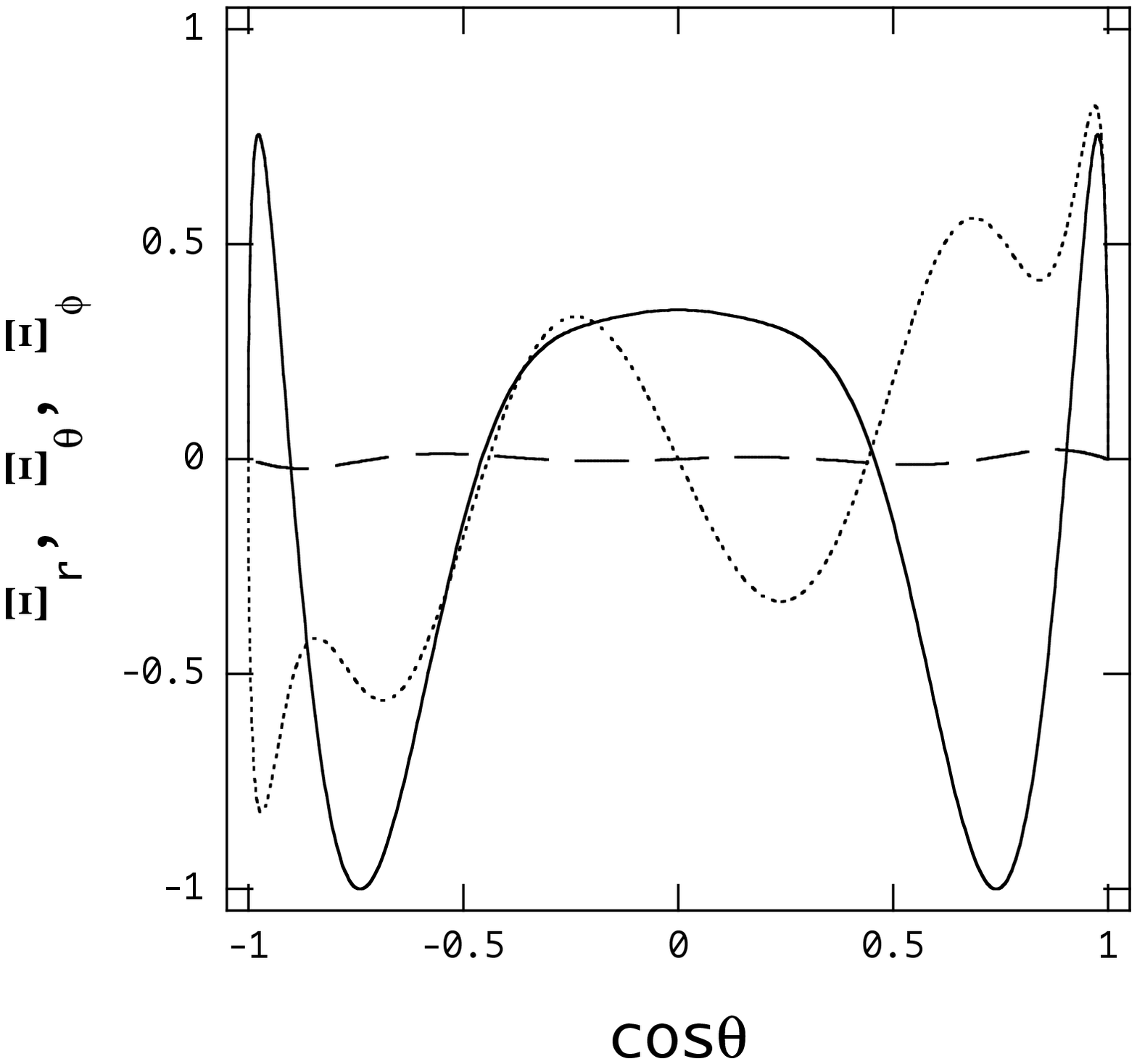}}
\caption{$\Xi_r$, $\Xi_\theta$, and $\Xi_\phi$ versus $\cos\theta$ for the $l'=m=2$ $r$-modes
calculated for a neutron star model composed of a fluid core, a solid crust, and a fluid ocean,
where the dashed, solid, and dotted curves are for $\Xi_r$, $\Xi_\theta$, and $\Xi_\phi$, respectively,
and the amplitudes are normalized by ${\rm max}(|\Xi_r|,|\Xi_\theta|,|\Xi_\phi|)$.
From the left to right panels, $\bar\Omega\equiv\Omega/\sqrt{GM/R^3}=0.1$, 0.2, and 0.3, respectively.
}
\end{figure}

Let us write as $\hat{\pmb{r}}_s=R\hat{\pmb{n}}_s$ the vector pointing from the stellar centre 
to the centre of a circular hot spot in the co-rotating frame, where
\be
\hat{\pmb{n}}_s=\sin\hat\theta_s\cos\hat\phi_s\hat{\pmb{i}}+\sin\hat\theta_s\sin\hat\phi_s\hat{\pmb{j}}+\cos\hat\theta_s\hat{\pmb{k}},
\ee
and $\hat\theta_s$ denotes the colatitude of the spot centre measured from the spin axis of the star, 
and $\hat{\pmb{i}}$,
$\hat{\pmb{j}}$, and $\hat{\pmb{k}}$ are the orthonormal vectors in the $\hat x$, $\hat y$, and $\hat z$ directions, respectively.
If the angular radius of the circular hot spot is equal to $\alpha$, we have for points $\hat{\pmb{r}}=R\hat{\pmb{n}}$ on the spot
\be
\hat{\pmb{n}}_s\cdot\hat{\pmb{n}}\ge\cos\alpha,
\ee
where $\hat{\pmb{n}}=\sin\hat\theta\cos\hat\phi\hat{\pmb{i}}+\sin\hat\theta\sin\hat\phi\hat{\pmb{j}}+
\cos\hat\theta\hat{\pmb{k}}$.
The outer boundary of the circular spot is given by
$\hat{\pmb{n}}_s\cdot\hat{\pmb{n}}_b=\cos\alpha$, that is,
\be
\sin\hat\theta_b\sin\hat\theta_s\cos\left(\hat\phi_b-\hat\phi_s\right)+\cos\hat\theta_b\cos\hat\theta_s=\cos\alpha,
\ee
where $\hat{\pmb{n}}_b=\sin\hat\theta_b\cos\hat\phi_b\hat{\pmb{i}}+\sin\hat\theta_b\sin\hat\phi_b\hat{\pmb{j}}
+\cos\hat\theta_b\hat{\pmb{k}}$.
If the hot spot is deformed by an $r$-mode having the displacement vector $\pmb{\xi}$, the outer boundary of the spot is approximately given by
\be
(\hat r_d,\hat\theta_d,\hat\phi_d)=(R,\hat\theta_b+\delta\theta,\hat\phi_b+\delta\phi)
\ee
where 
\be
\delta\theta=\xi_\theta(R,\hat\theta_b,\hat\phi_b)/R=\Xi_\theta(\hat\theta_b)\cos(m\hat\phi_b+\omega t)
\ee
and 
\be
\delta\phi=\xi_\phi(R,\hat\theta_b,\hat\phi_b)/(R\sin\hat\theta_b)=-\Xi_\phi(\hat\theta_b)\sin(m\hat\phi_b+\omega t)/\sin\hat\theta_b
\ee
where $|\pmb{\xi}/R|\ll1$ is assumed.

If using $\hat{\pmb{n}}$ we define a vector $\hat{\pmb{n}}_\perp$ that is perpendicular to $\hat{\pmb{n}}_s$ as
\be
\hat{\pmb{n}}_\perp=\hat{\pmb{n}}-\left(\hat{\pmb{n}}\cdot\hat{\pmb{n}}_s\right)\hat{\pmb{n}}_s\equiv-Y\hat{\pmb{e}}_\theta^s+X\hat{\pmb{e}}_\phi^s,
\ee
we have
\be
X=\sin\hat\theta\sin\left(\hat\phi-\hat\phi_s\right),
\quad
Y=-\cos\hat\theta_s\sin\hat\theta\cos\left(\hat\phi-\hat\phi_s\right)+\sin\hat\theta_s\cos\hat\theta,
\ee
where $\hat{\pmb{e}}_\theta^s$ and $\hat{\pmb{e}}_\phi^s$ are the orthonormal vectors in the $\hat\theta$ and $\hat\phi$
directions and are perpendicular to $\hat{\pmb{e}}_r^s=\hat{\pmb{n}}_s$.
For $\hat{\pmb{n}}=\hat{\pmb{n}}_b$ that satisfies equation (18), we obtain
\be
X_b^2+Y_b^2=1-\cos^2\alpha,
\ee
where
\be
X_b=\sin\hat\theta_b\sin\left(\hat\phi_b-\hat\phi_s\right),
\quad
Y_b=-\cos\hat\theta_s\sin\hat\theta_b\cos\left(\hat\phi_b-\hat\phi_s\right)+\sin\hat\theta_s\cos\hat\theta_b.
\ee
For $\hat{\pmb{n}}=\hat{\pmb{n}}_d$, on the other hand, we have
\be
X_d=\sin(\hat\theta_b+\delta\theta)\sin\left(\hat\phi_b+\delta\phi-\hat\phi_s\right),
\quad
Y_d=-\cos\hat\theta_s\sin(\hat\theta_b+\delta\theta)\cos\left(\hat\phi_b+\delta\phi-\hat\phi_s\right)
+\sin\hat\theta_s\cos(\hat\theta_b+\delta\phi).
\ee
A plot of $(X_d,Y_d)$ in a plane may be regarded as a projection of the outer boundary of the deformed spot
onto a plane perpendicular to the vector $\hat{\pmb{n}}_s$.
Examples of the plots $(X_b,Y_b)$ and $(X_d,Y_d)$ are given in Figure 2 for $\bar\Omega=0.1$ and in Figure 3 for $\bar\Omega=0.3$, 
where for the spot of $\alpha=20^\circ$ we have assumed 
$\theta_s=10^\circ$, $20^\circ$, and $30^\circ$, from the left to right panels in each figure,
and $A=0.1$ for the amplitudes of the functions $\Xi_\theta$ and $\Xi_\phi$ of the $l'=m=2$ $r$-modes given in Figure 1.
%
Note that we have assumed for simplicity
that the oscillation frequency of the $l'=m=2$ $r$-mode
is exactly equal to $\omega=2m\Omega/[l'(l'+1)]=2\Omega/3$ in the co-rotating frame.
The deviation of the $r$-mode frequency from the analytic formula $2m\Omega/[l'(l'+1)]$
for neutron stars depends on various parameters such as
the spin frequency, the solid crust thickness, the equation of state, the thermal 
stratification in the core,
the general relativistic effects and so on, and since it is not our main concern here to 
investigate how the light curves depend on the frequency deviation, we have simply assumed
$\omega=2m\Omega/[l'(l'+1)]$ for the modes.
The figures show periodic deformation of the spot shape, caused by the $l'=m=2$ $r$-mode, as seen in the
co-rotating frame of the star.
As shown by the middle panels of the figures, for $\alpha=20^\circ$ and $\theta_s=20^\circ$ the outer boundary of the spot
touches the pole of $\hat\theta=0$, at which $\Xi_\theta$ and $\Xi_\phi$ vanish.
Since the amplitudes of the functions $\Xi_\theta$ and $\Xi_\phi$ show a sharp increase around the poles
for $\bar\Omega=0.3$, the outer boundary of the spot is largely deformed when $\theta_s\sim\alpha$.

\begin{figure}
\resizebox{0.33\columnwidth}{!}{
\includegraphics{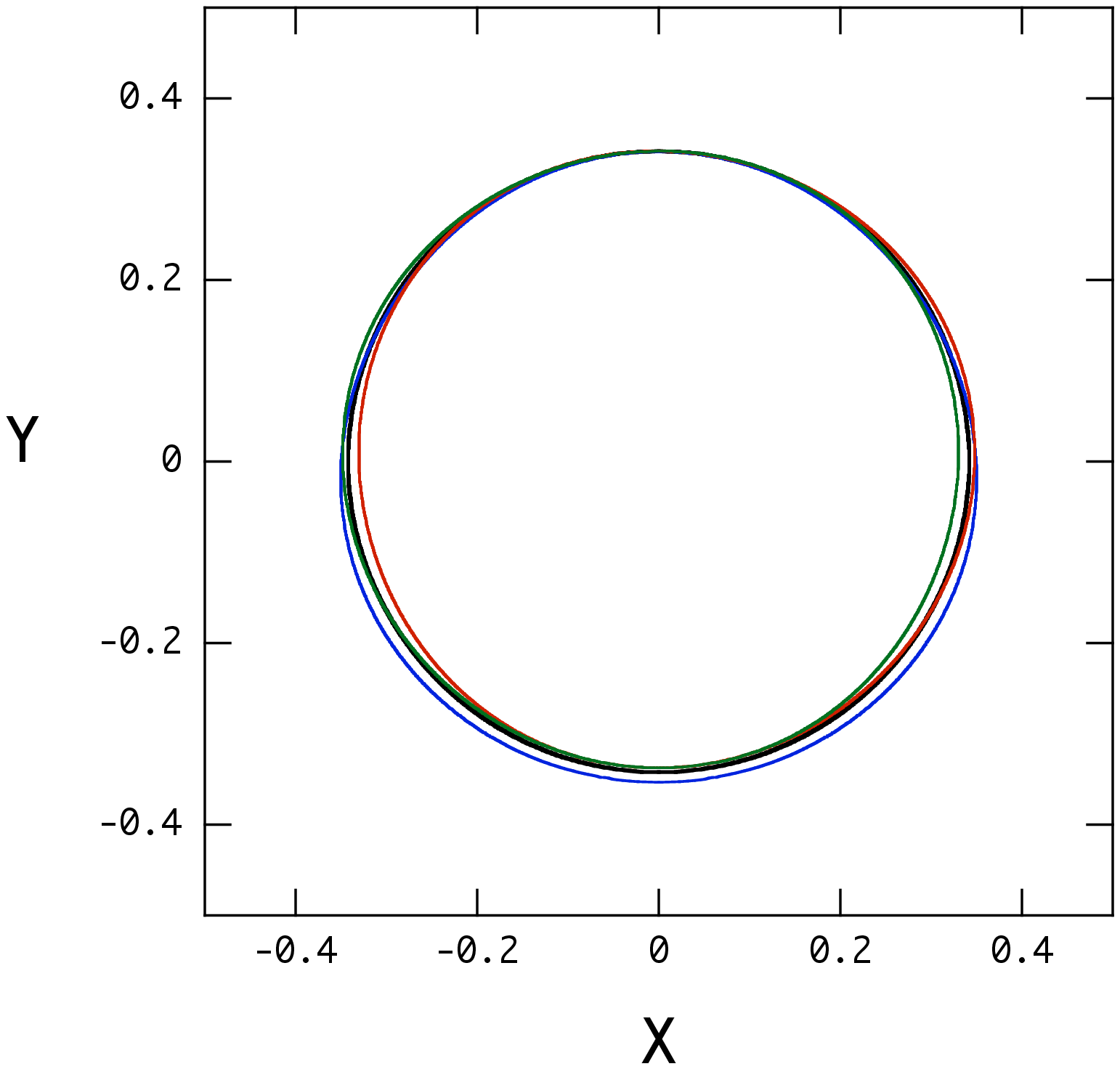}}
\resizebox{0.33\columnwidth}{!}{
\includegraphics{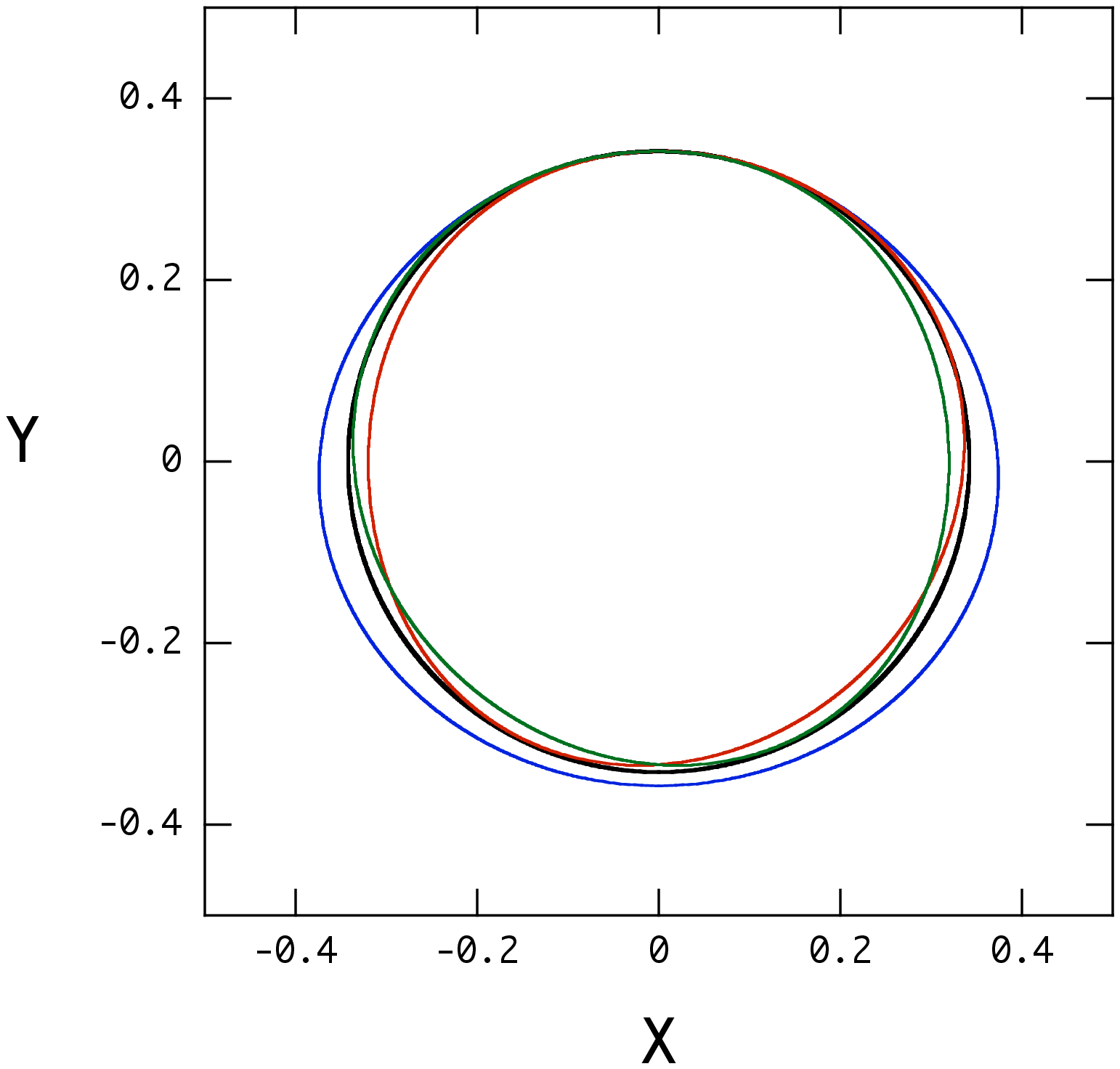}}
\resizebox{0.33\columnwidth}{!}{
\includegraphics{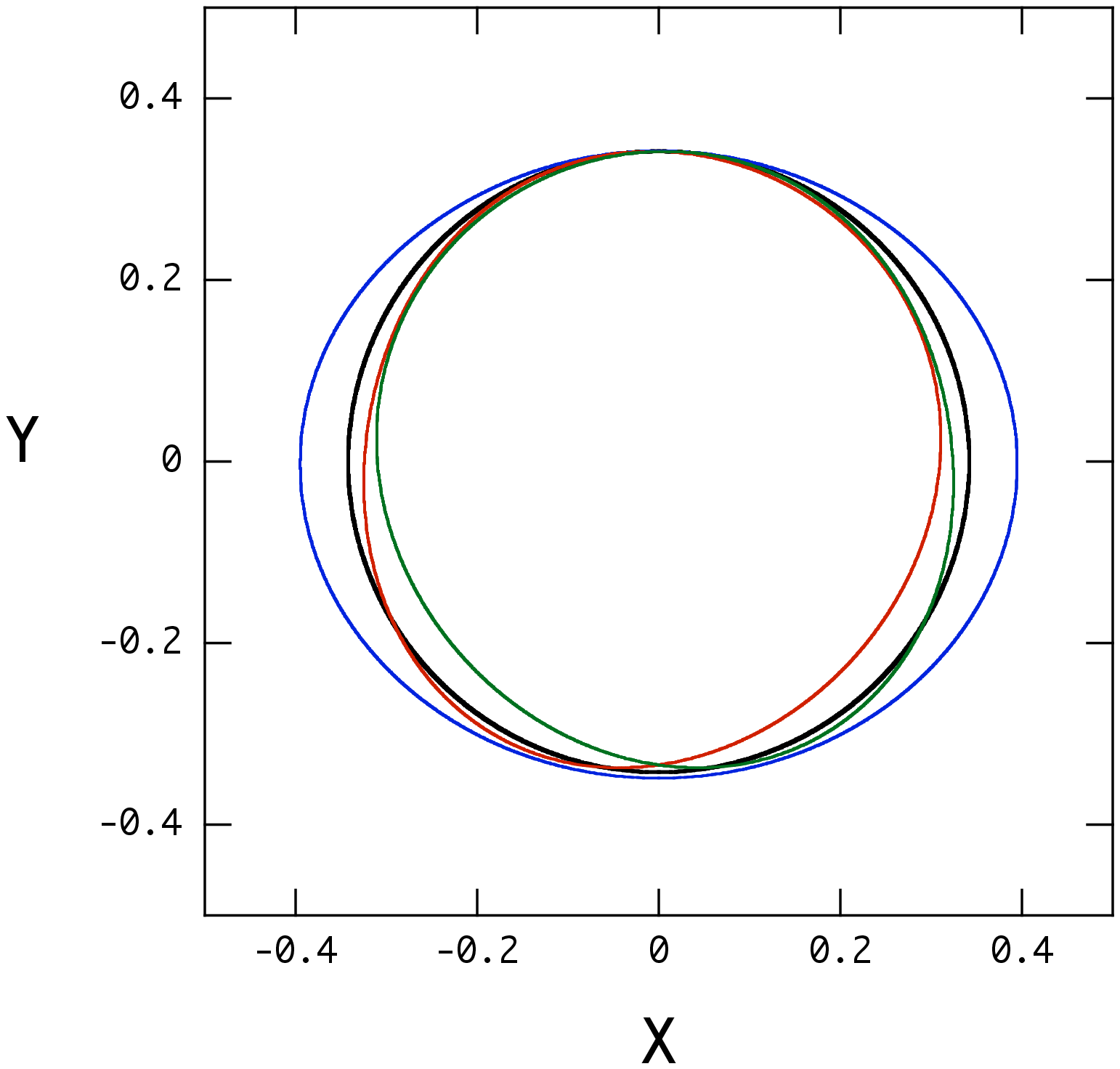}}
\caption{Plots of $(X_b,Y_b)$ and $(X_d,Y_d)$ for the spot of $\alpha=20^\circ$ computed using the functions $\Xi_\theta$ and $\Xi_\phi$ for the $l'=m=2$ $r$-mode at $\bar\Omega=0.1$,
where $\theta_s=10^\circ$, $20^\circ$ and $30^\circ$, from the left to right panels, and
we assume $\omega=2\Omega/3$ and $A=0.1$.
Here, the black curve is for the non-perturbed spot $(X_b,Y_b)$ and the blue, red, and green curves are
for the perturbed spot $(X_d,Y_d)$ at $\Omega t/(2\pi)=0, ~1, ~2$, respectively.}
\end{figure}

\begin{figure}
\resizebox{0.33\columnwidth}{!}{
\includegraphics{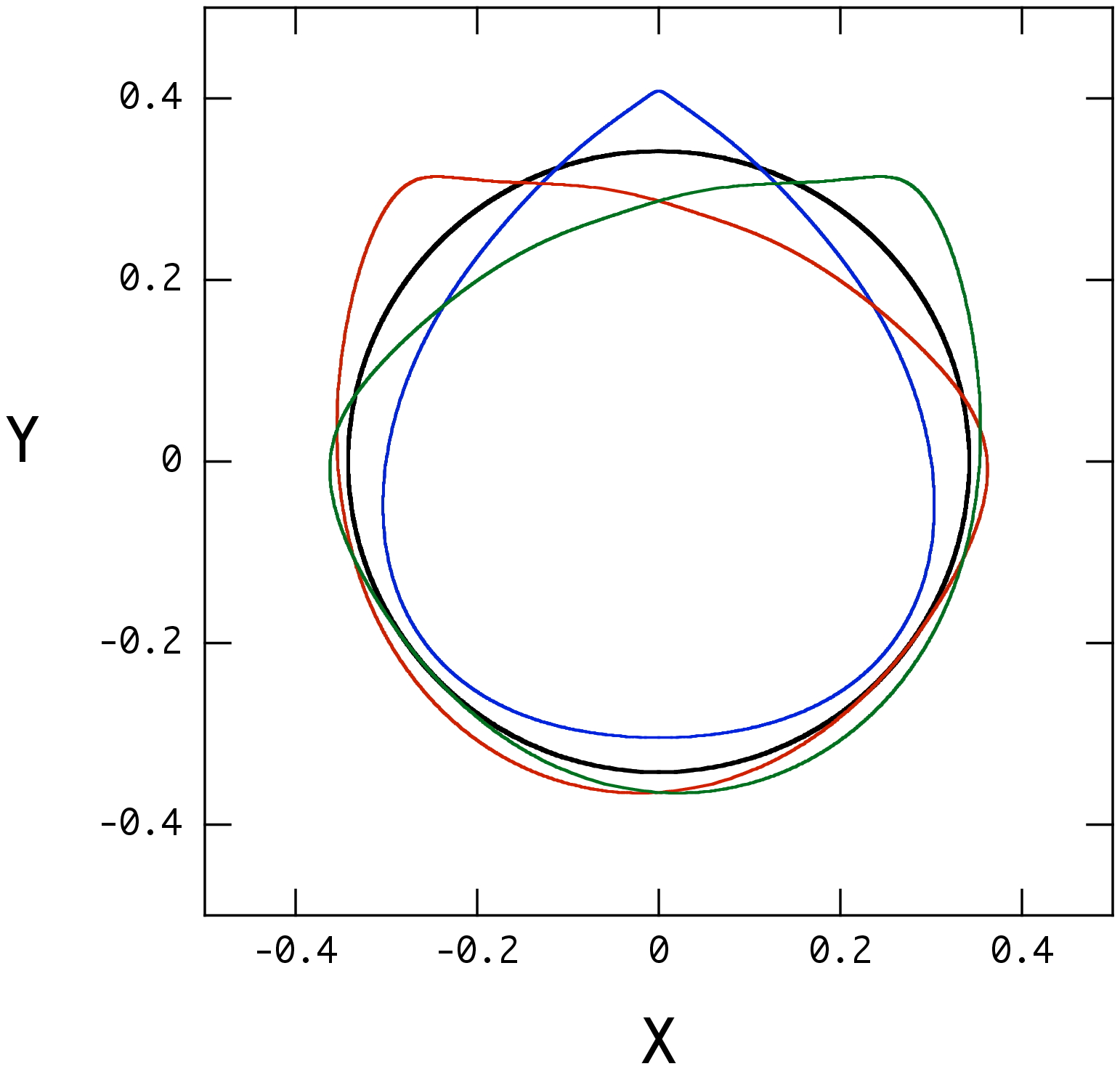}}
\resizebox{0.33\columnwidth}{!}{
\includegraphics{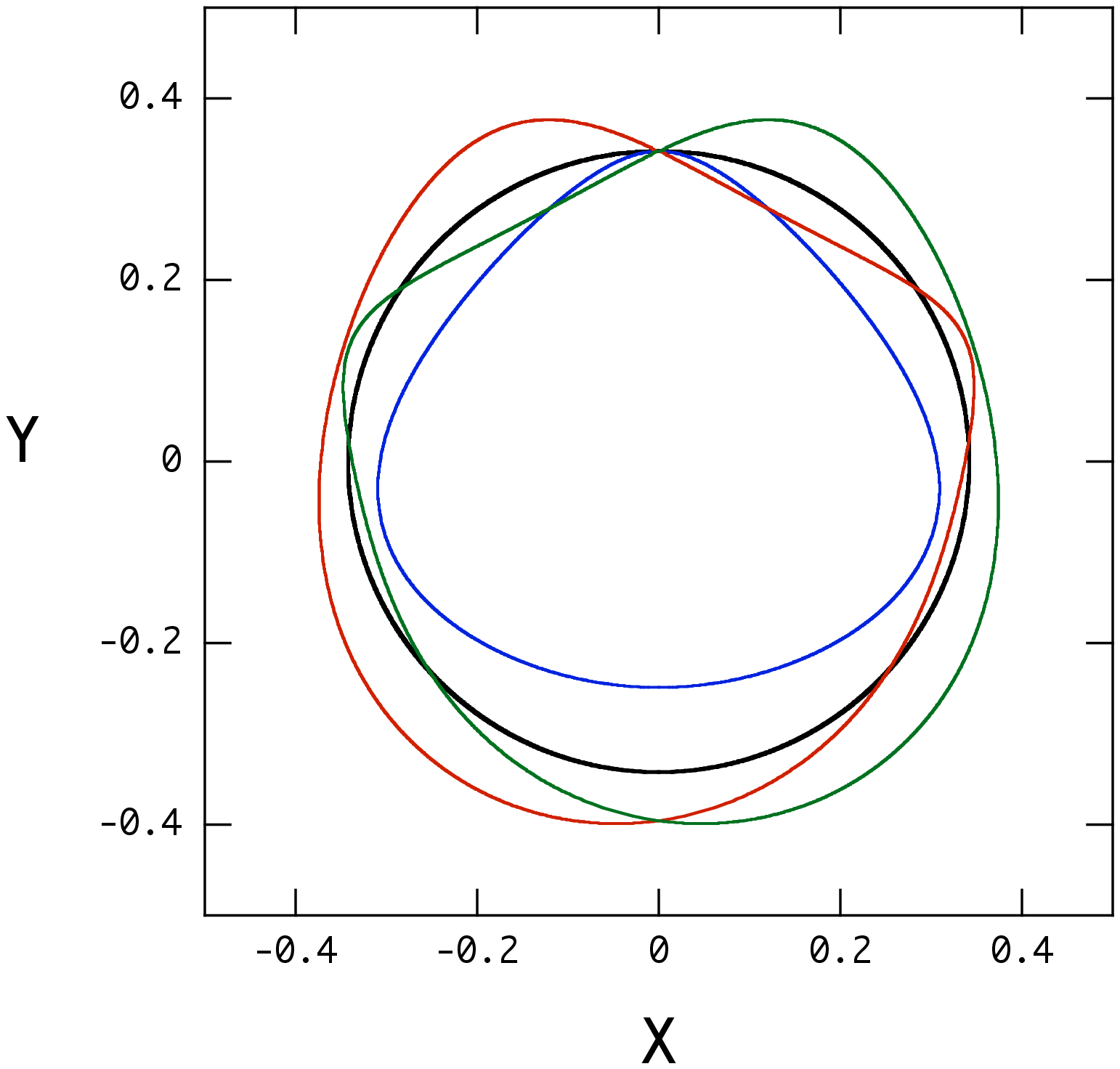}}
\resizebox{0.33\columnwidth}{!}{
\includegraphics{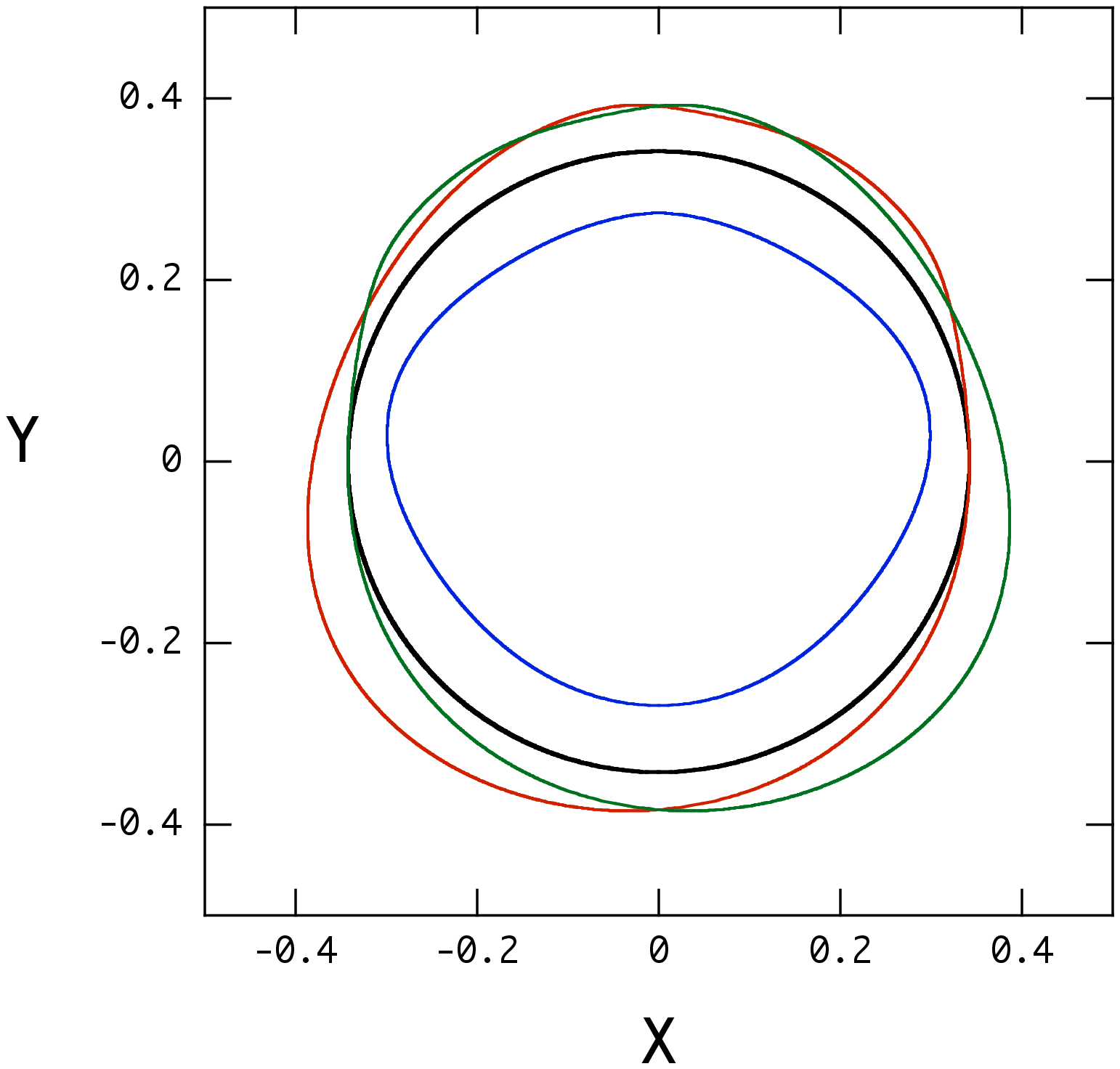}}
\caption{Same as Figure 2 but for $\bar\Omega=0.3$.}
\end{figure}

\section{numerical results}

Examples of light curves produced by a hot spot of $\alpha=20^\circ$
are plotted as a function of $\Omega t/(2\pi)$ in Figure 4 for $\bar\Omega=0.1$ and in Figure 5 for $\bar\Omega=0.3$, 
where $\theta_s=10^\circ$ for panel (a) and $30^\circ$ for panel (b), and we have assumed $M=1.4M_\odot$, $R=10^6$cm 
for the neutron star, and $\omega=2\Omega/3$ and $A=0.01$ for the $l'=m=2$ $r$-modes.
Here, we have also assumed that $\hat I_0$ is constant.
The figures show that the amplitudes of $\delta F=(F-F_m)/F_m$ increase when
the angular distance $\theta_s$ of the spot centre or the inclination angle $i$ increases, 
where $F_m$ is the mean flux.
An increase in the spin frequency also tends to increase the amplitudes of the variations $\delta F$ through the Doppler factor
particularly for large values of $\theta_s$ and $i$.
For a oscillation amplitude $A=0.01$, we can clearly see the $r$-modes cause periodic modulations of the amplitudes of $\delta F$.

\begin{figure}
\resizebox{0.5\columnwidth}{!}{
\includegraphics{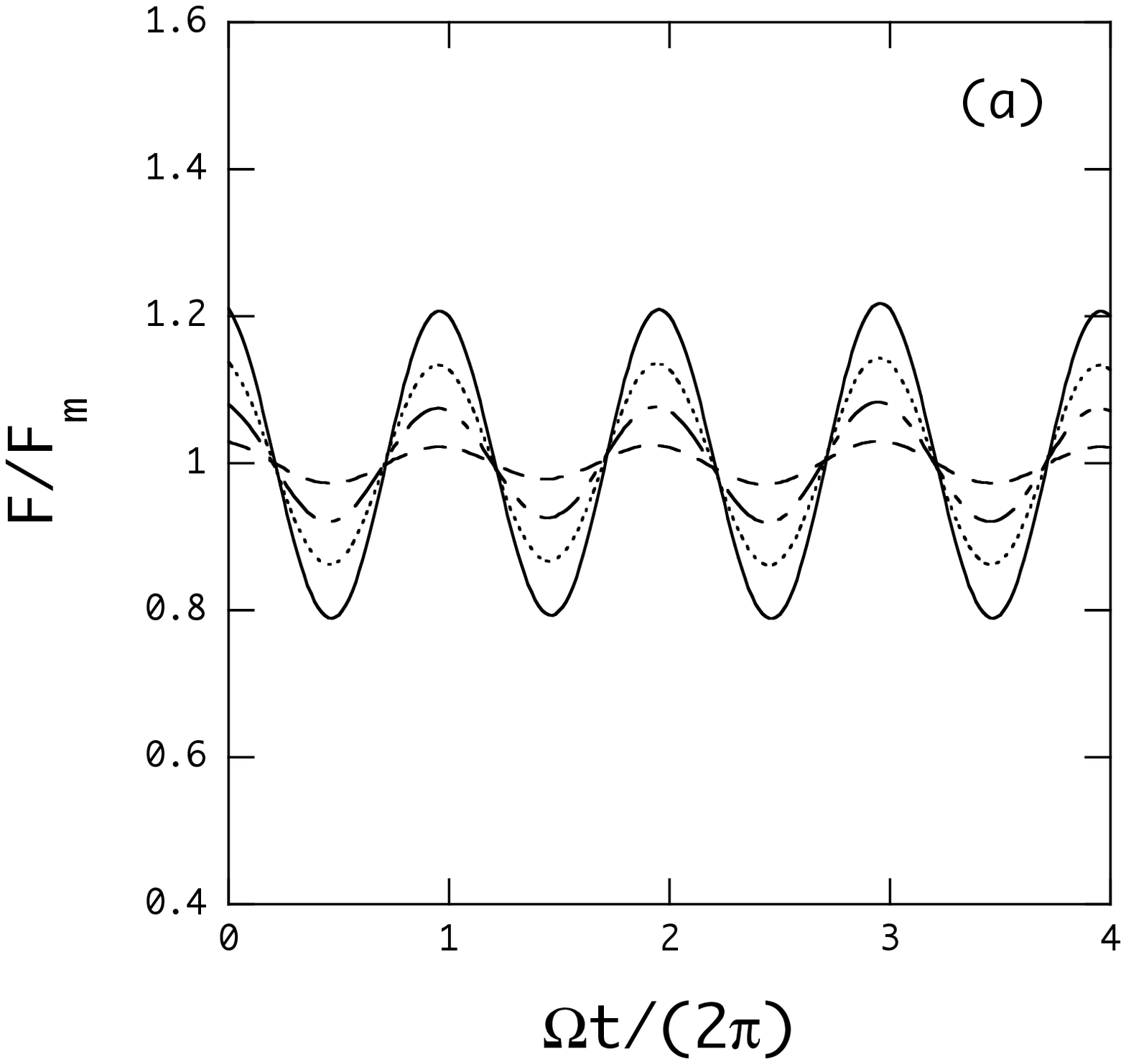}}
\resizebox{0.5\columnwidth}{!}{
\includegraphics{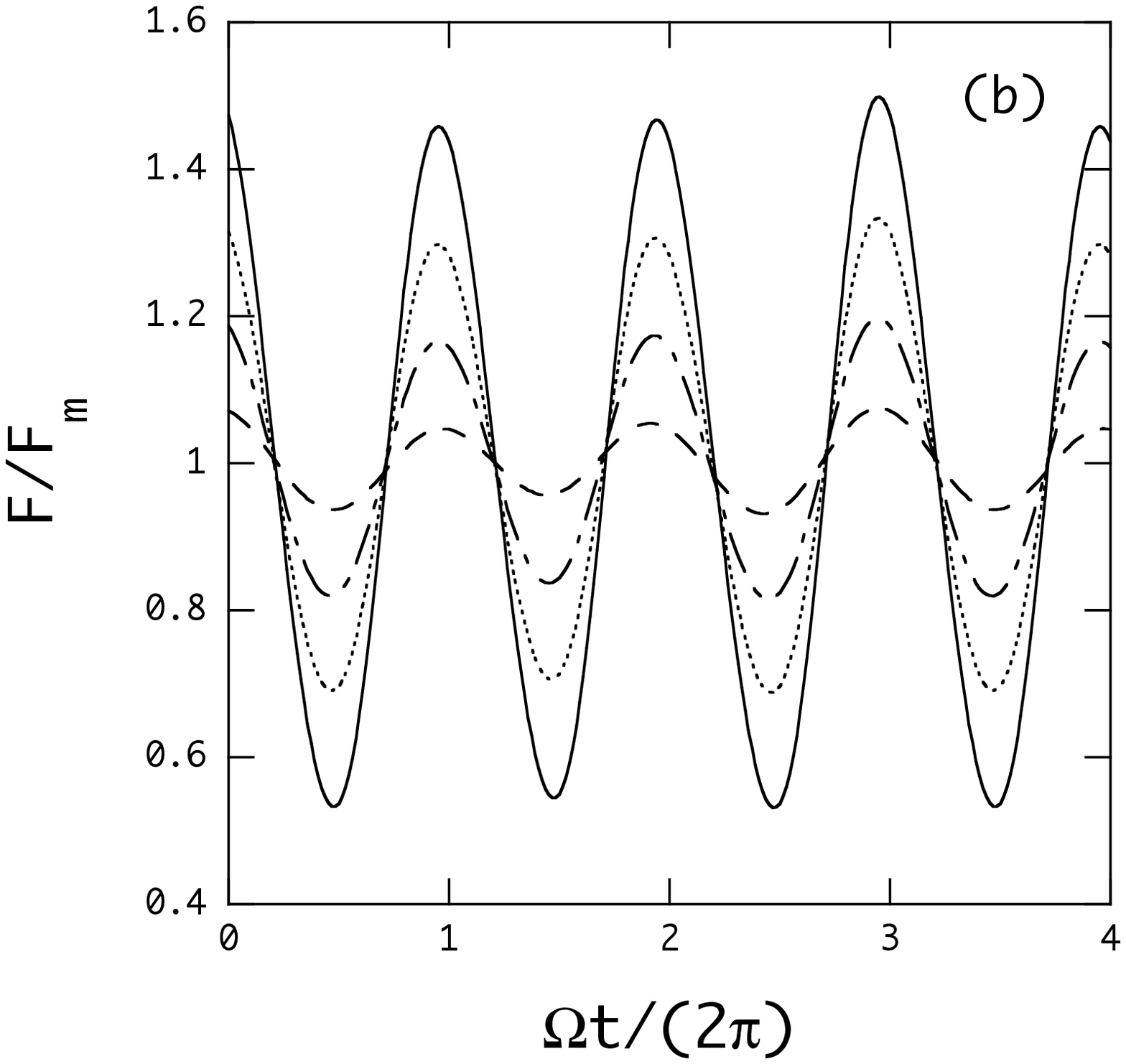}}
\caption{Light curves $F/F_m$ produced by a hot spot of $\alpha=20^\circ$ for $\bar\Omega=0.1$
as functions of $\Omega t/(2\pi)$ for $\theta_s=10^\circ$ (panel a) and $30^\circ$ (panel b), 
where $F_m$ is the mean flux, and the dashed, dash-dotted, dotted, and solid lines are
for the inclination angle $i=10^\circ,~30^\circ,~50^\circ$, and $70^\circ$, respectively.
Here, we assume $M=1.4M_\odot$, $R=10^6$cm for the neutron star, and $\omega=2\Omega/3$ and $A=0.01$ for the $l'=m=2$ $r$-mode.}
\end{figure}

\begin{figure}
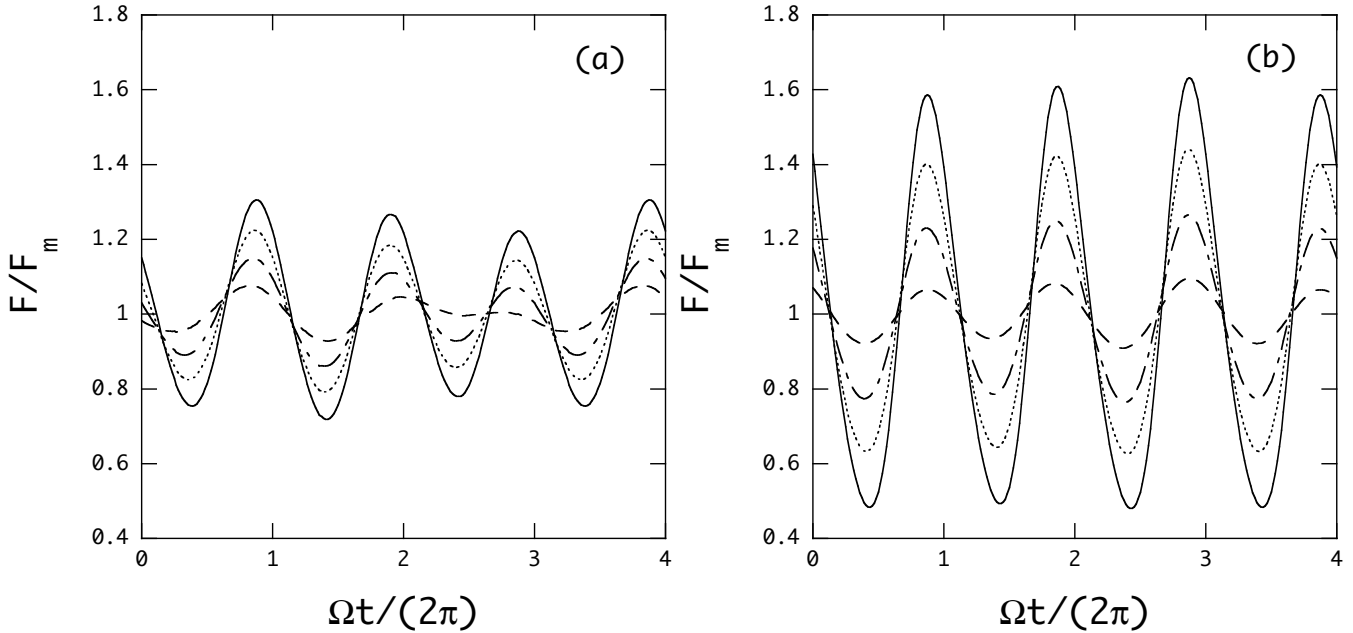

\resizebox{0.5\columnwidth}{!}{
\includegraphics{f5a.epsi}}
\resizebox{0.5\columnwidth}{!}{
\includegraphics{f5b.epsi}}
\caption{Same as Figure 4 but for $\bar\Omega=0.3$.}
\end{figure}

Using light curves $F(t)$,
we calculate the discrete Fourier transform $a_j$ ($j=-N/2,\cdots,N/2-1$) defined by
\be
a_j=\sum_{k=0}^{N-1}F(t_k)\exp\left(2\pi{\rm i}f_jt_k\right),
\ee
where $N$ is the total number of sampling points in the time-span $\Delta T$, $t_k=k\Delta T/N$,
$f_j=j/\Delta T$, and $|a_j|=|a_{-j}|$ for a real function $F(t)$, and $k$ and $j$ are integers.
For light curve calculations, we use $N=N_1N_2$ with $N_1=2^6$ and $N_2=2^5$ so that we have the 
Nyquist frequency $\nu_{Ny}=/(2\delta t)$ with $\delta t=P_s/N_1$
and the time-span $\Delta T=P_sN_2$ where $P_s=2\pi/\Omega$ is the spin period.
In Figure 6 plotted are the fractional Fourier amplitudes $a_j/a_0$, which is proportional to 
the fractional rms, as functions of $\sigma_j/\Omega$ with $\sigma_j\equiv 2\pi f_j$
for light curves calculated assuming
$M=1.4M_\odot$, $R=10^6$cm, and 
$\bar\Omega=0.3$.
The dominant peak of $a_j/a_0$ appears at $\sigma_j=\Omega$
due to the spin frequency $\Omega$ of the star, which we may call the fundamental, and there also appear weaker peaks 
at $\sigma_j=2\Omega$ (first overtone) and $3\Omega$ (second overtone).
Because of the periodic modulations caused by the $l'=m=2$ $r$-mode having the frequency $\omega=2m\Omega/[l'(l'+1)]=2\Omega/3$ in the co-rotating frame, we also have a noticeable peak at $\sigma_j=2\Omega/3$.
Note that a distant observer 
will detect the $r$-mode frequency measured in the co-rotating frame,
instead of that in an inertial frame, because we are seeing waves restricted to the spot co-moving with the star.  
Although the peak at $\sigma_j=2\Omega/3$ is almost insensitive to the inclination angle $i$,
the peak at $\sigma_j=\Omega$ becomes higher as $i$ or $\theta_s$ increases (e.g., Lamb et al 2009).
It is reasonable that the peak at $\sigma_j=2\Omega/3$ is approximately proportional to the amplitude parameter $A$ 
but the peak at $\sigma_j=\Omega$ is insensitive to it.
We also note that much weaker peaks, which seem to be proportional to the parameter $A$, are found at $\sigma_j=k\Omega\pm j2\Omega/3$ with $k$ and $j$ being integers as a result of nonlinear couplings between the frequencies $k\Omega$ and
$j2\Omega/3$.

The fractional amplitudes $a_j/a_0$ 
are plotted versus $\theta_s$ in Figure 7 for $\bar\Omega=0.1$ and in Figure 8 for $\bar\Omega=0.3$.
In general, the peaks at $\sigma_j=\Omega$ and $2\Omega$ increase their height with increasing $\theta_s$.
On the other hand, the peak at $\sigma_j=2\Omega/3$ shows rather complicated behavior with increasing $\theta_s$
depending on the parameter $\alpha$ and the functions $\Xi_\theta(\hat\theta)$ and $\Xi_\phi(\hat\theta)$, but
as $\theta_s\rightarrow 0$ its height simply increases to become even higher than the peak at $\sigma_j=2\Omega$, 
because in this limit 
the periodic deformation of the spot shape is the only cause for any periodicities.
For the amplitude parameter $A=10^{-3}$, for example, the amplitude $a_j/a_0$ at $\sigma_j=2\Omega/3$ stays less than $10^{-3}$ when
$\theta_s\gtsim 10^\circ$ for both cases of $\bar\Omega=0.1$ and 0.3, but it becomes as large as
$\sim0.01$ for small values of $\theta_s$.
We also note that although the peaks at $\sigma_j=\Omega$ and $2\Omega$ are dependent on the angle $i$, 
the peak at $\sigma_j=2\Omega/3$ is almost insensitive to $i$ because no effects of the
velocity field due to the $r$-modes are included.
Figure 9 gives plots of $(X_b,Y_b)$ and $(X_d,Y_d)$ for the case of $\bar\Omega=0.3$, $\alpha=20^\circ$, and 
$\theta_s=50^\circ$ corresponding to the deep dip found at $\theta_s=50^\circ$ in the panel (a) of Figure 8.
We find that differences between the spot deformations in different phases $\Omega t$
are rather small compared with those found in the plots of $(X_b,Y_b)$ and $(X_d,Y_d)$ in Figure 3, leading to
the dip in $a_j/a_0$ associated with $\sigma_j=2\Omega/3$.
We have also examined the dependence of the amplitudes $a_j/a_0$ at $\sigma_j=2\Omega/3$ on
the compactness parameter $r_g/R$ and have found that the amplitudes are only weakly dependent 
on the parameter.

\begin{figure}
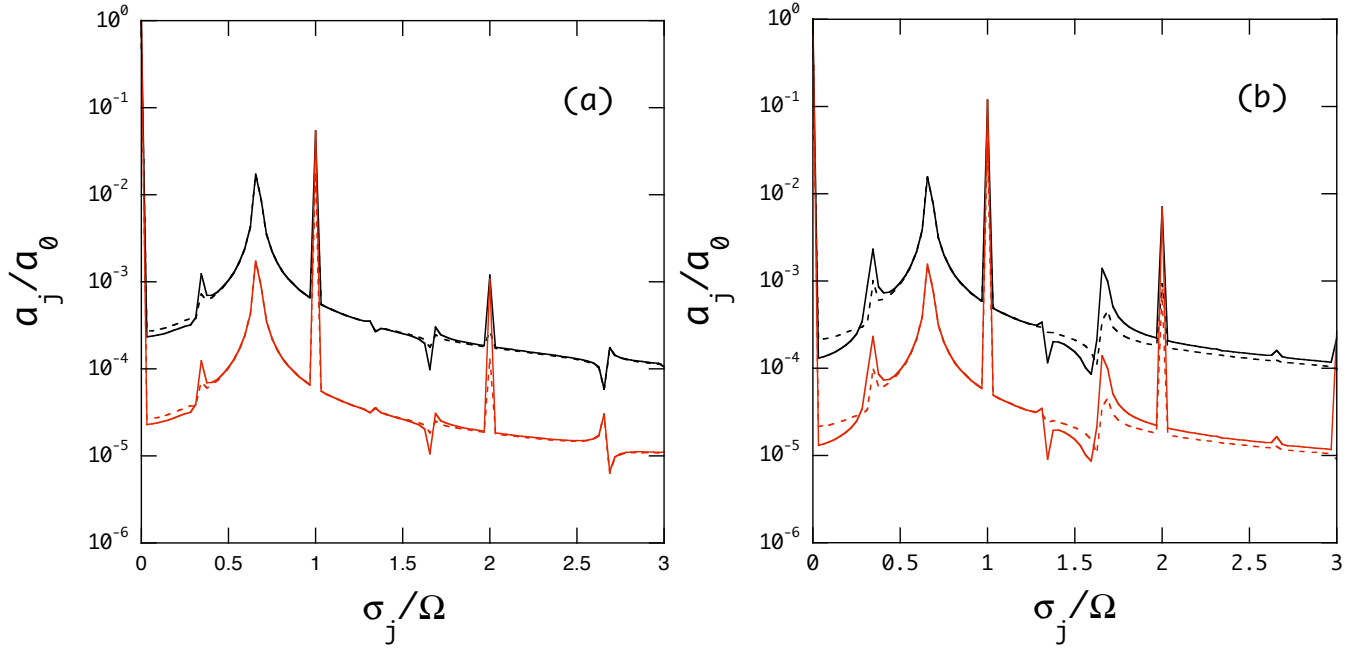

\resizebox{0.5\columnwidth}{!}{
\includegraphics{f6a.epsi}}
\resizebox{0.5\columnwidth}{!}{
\includegraphics{f6b.epsi}}
\caption{Normalized Fourier amplitudes $a_j/a_0$ as functions of $\sigma_j/\Omega$ calculated for light curves produced by 
a hot spot of $\alpha=20^\circ$ for $\theta_s=10^\circ$ (panel a) and $30^\circ$ (panel b), 
where $\bar\Omega=0.3$, and the black and red lines are for $A=0.01$ and $A=0.001$, respectively, and 
the solid and dotted lines are for the inclination angle $i=30^\circ$ and $10^\circ$, respectively.
Here, we assume $M=1.4M_\odot$ and $R=10^6$cm.}
\end{figure}

\begin{figure}
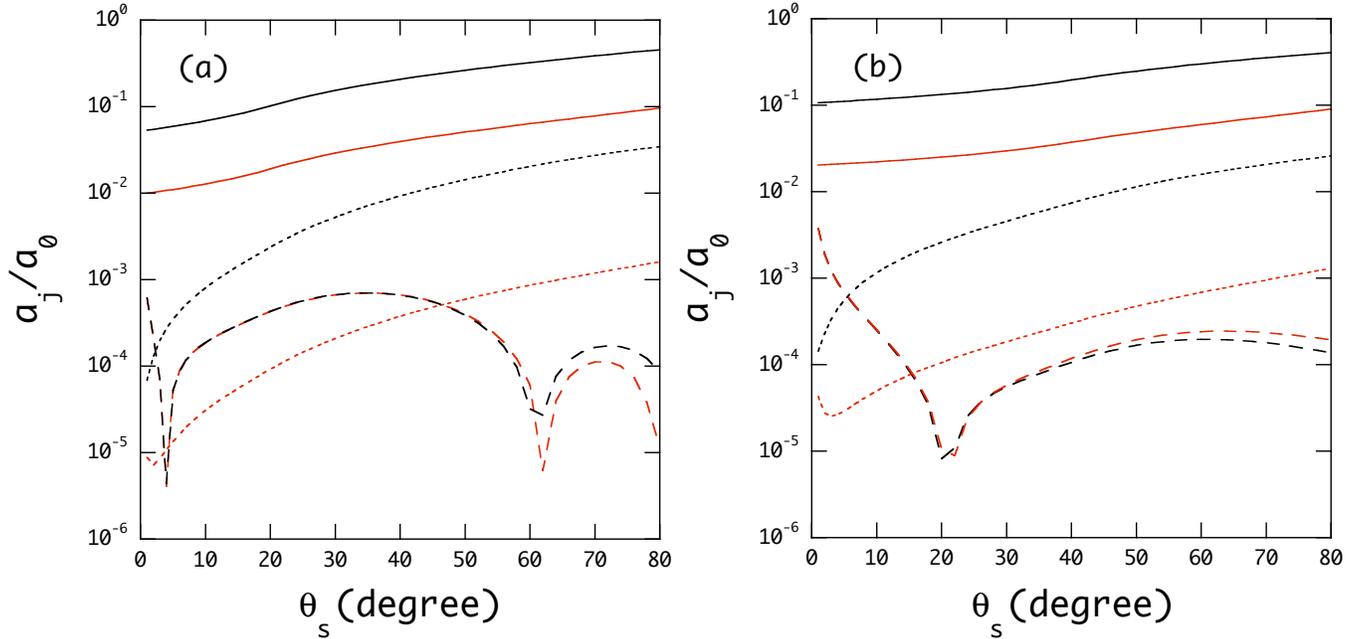

\resizebox{0.5\columnwidth}{!}{
\includegraphics{f7a.epsi}}
\resizebox{0.5\columnwidth}{!}{
\includegraphics{f7b.epsi}}
\caption{Normalized Fourier amplitudes $a_j/a_0$ as functions of the angular distance $\theta_s$ of
the spot centre from the rotation axis of the star for $\alpha=20^\circ$ (panel a) and
$\alpha=40^\circ$ (panel b), where the black and red lines are for $i=50^\circ$ and $i=10^\circ$,
respectively, and 
the solid, dotted, and dashed lines are for the fundamental $\sigma_j=\Omega$, the
first overtone $\sigma_j=2\Omega$, and the $l'=m=2$ $r$-mode $\sigma_j=2\Omega/3$, respectively.
Here, we assume $M=1.4M_\odot$, $R=10^6$cm, and $\bar\Omega=0.1$ for the neutron star, and $A=0.001$ for the mode.}
\end{figure}

\begin{figure}
\resizebox{0.5\columnwidth}{!}{
\includegraphics{f8a.epsi}}
\resizebox{0.5\columnwidth}{!}{
\includegraphics{f8b.epsi}}
\caption{Same as Figure 7 but for $\bar\Omega=0.3$}
\end{figure}

\begin{figure}
\resizebox{0.33\columnwidth}{!}{
\includegraphics{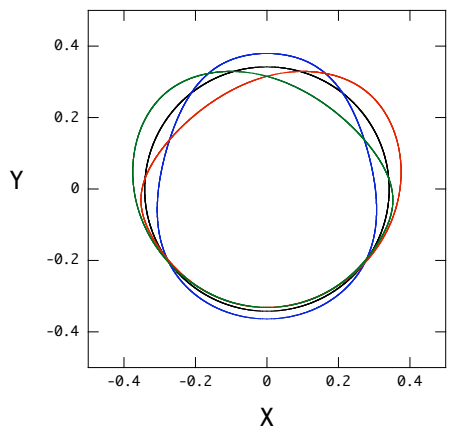}}
\caption{Plots of $(X_b,Y_b)$ and $(X_d,Y_d)$ for the spot of $\alpha=20^\circ$ and $\theta_s=50^\circ$,
where we have used the functions $\Xi_\theta$ and
$\Xi_\phi$ computed for the $l'=m=2$ $r$-mode at $\bar\Omega=0.3$ and have assumed $\omega=2\Omega/3$ and $A=0.1$
for the plots.
Here, the black curve is for the non-perturbed spot $(X_b,Y_b)$ and the blue, red, and green curves are
for the perturbed spot $(X_d,Y_d)$ for $\Omega t/(2\pi)=0, ~1, ~2$, respectively.}
\end{figure}


\section{conclusion}

We have calculated light curves produced by a hot spot of a rapidly rotating neutron star,
assuming the hot spot is periodically disturbed by the horizontal displacement field
of the $l'=m=2$ core $r$-mode, which is assumed to be excited by emitting gravitational waves.
To calculate light curves, 
we have taken account of relativistic effects such as the Doppler boost due to the rapid rotation and 
light bending assuming the Schwarzschild metric around the star.
We have also assumed that the oscillation frequency of the $l'=m=2$ core $r$-mode
is exactly equal to $\omega=2\Omega/3$ in the co-rotating frame of the star.
It is found that
a distant observer will detect in the light curves a periodicity due to the $l'=m=2$ core $r$-mode and that the
observed frequency will be $2\Omega/3$, the frequency of the mode defined in the
co-rotating frame of the star, instead of the frequency $4\Omega/3$ defined in an inertial frame for the mode.
The fractional Fourier amplitude $a_j/a_0$ at $\sigma_j=2\Omega/3$ in light curves is approximately proportional to the amplitude parameter $A$ which parametrizes the mode amplitude such that ${\rm max}\left(|\Xi_\theta(\hat\theta)|,|\Xi_\phi(\hat\theta)|\right)=A$ for $0\le\hat\theta\le\pi$.
We find that besides the parameter $A$ the amplitude $a_j/a_0$ at $\sigma_j=2\Omega/3$ depends on the parameter $\theta_s$,
but it is almost insensitive to the parameters $i$ and $r_g/R$.
The reason for the insensitivity of $a_j/a_0$ to the parameters may be that
no effects of the velocity field due to the $r$-mode on light curves are taken into account of in the calculations
and the periodic deformation of the spot shape is restricted to a small area on the surface.
For $A=0.001$ to $0.01$, the amplitude $a_j/a_0$ at $\sigma_j=2\Omega/3$ will be $\sim0.001$ to $\sim0.01$ and
becomes comparable to or 
even greater than that of the first overtone at $\sigma_j=2\Omega$, particularly for small values of $\theta_s$.

If we write the oscillation frequency of the $r$-modes observed in the co-rotating frame of the star as
\be
{\omega/\Omega}=\kappa_0+\kappa_2\bar\Omega^2+O\left(\bar\Omega^4\right),
\ee 
the coefficient $\kappa_0$ for the $r$-modes with $l'$ and $m$ is simply given by
\be
\kappa_0=2m/ \left[l^\prime\left(l^\prime+1\right)\right],
\ee
and the coefficient $\kappa_2$ depends on the physical properties of neutron stars such as the equation of state and the deviation from
the isentropic stratification in the core (e.g., Yoshida \& Lee 2000a,b).
Since the neutron star core is nearly isentropic such that $N^2\sim0$ with $N$ being the Brunt-V\"ais\"al\"a frequency, we only
have to consider the $l^\prime=m$ $r$-modes, for which we have 
$\omega\approx\kappa_0\Omega=2\Omega/\left(m+1\right)$, and we obtain
the frequency $\omega\approx 2\Omega/3$ for $m=2$
in the co-rotating frame of the star.
Although no detection of periodicities $\omega\approx2\Omega/3$ associated with the $l'=m=2$ core $r$-mode
has so far been reported,
any detection of periodicities caused by the $l^\prime=m$ core $r$-modes in the X-ray millisecond pulsation
makes it possible to use the frequency deviation given by 
$\Delta\bar\omega\equiv\bar\omega-\kappa_0\bar\Omega\approx\kappa_2\bar\Omega^3$ to derive
information about the equations of state and the thermal stratification in the core.

Employing spcae-time metric numerically computed for rotating neutron stars in stead of the Schwarzschild metric,
Cadeau et al (2005, 2007) have discussed the frame dragging effects on 
the light curves produced by a small hot spot on the surface of rapidly rotating neutron stars.
Cadeau et al (2007) have concluded that in most cases
the differences in the light curves between the choice of
metric are smaller than the differences caused by the spherical or oblate
shape of the initial emission surface.
We think this is also the case for light curves produced by a finite hot spot periodically modified
by the $r$-mode.
We may also guess that the effects of the rotational deformation of neutron stars on the frequency and surface wave-pattern
of the $r$-mode, for example, could be more significant than the frame dragging effects for light curves.
We believe that at the current stage of investigation the frame dragging may be regarded as one of the effects
that become discernible only after very accurate light curve determination becomes possible observationally and theoretically.

To use a surface hot spot of a rapidly rotating neutron star as a probe into core $r$-modes, 
the modes must penetrate the solid crust to have sufficient amplitudes at the surface.
The detectability of periodicities due to the core $r$-modes in light curves therefore may depend on, 
for example, the thickness of the solid crust,
the property of the surface fluid ocean, and the strength of the magnetic field and so on.
In this paper, although we have used the functions $\Xi_\theta$ and $\Xi_\phi$ computed for
a low mass, cold neutron star model with a thick solid crust, which makes the frequency spectrum simple, 
it is desirable to use the functions computed for more massive neutron stars as well, which may
have a thin solid crust and hence have a different oscillation frequency spectrum from that for
a cold low mass neutron star.
For accretion powered neutron stars,
it is also desirable to use mass accreting neutron star models with a hot fluid ocean and solid crust, with which we can examine
how the amplitudes of core $r$-modes at the surface depends on the physical properties of
the fluid ocean and solid crust.
The existence of a magnetic field is another important factor we have to consider,
particularly to determine the wave patterns at the surface.
Although the effects of a magnetic field on the displacement vector $\pmb{\xi}$ of 
core $r$-modes have been examined only for
a dipole field whose axis aligns with the spin axis, we need to examine how an oblique dipole
field and a field different from a dipole one change the surface wave patterns (e.g., Heng \& Spitkovsky 2009).

Applying a weak nonlinear theory, Arras et al (2003) have estimated the saturation level
for the $r$-mode energy $E_{\rm r-mode}$ by considering nonlinear transfer of energy to
the sea of stellar inertial oscillation modes of rotating stars with negligible buoyancy
and elastic restoring force and without magnetic field.
For the saturation energy in the strong driving limit of the $r$-mode, they obtained an estimate given by
$E_{\rm r-mode}/(0.5MR^2\Omega^2)\simeq10^{-6}(\nu_{\rm spin}/10^3{\rm Hz})^5$, which may amount to
the mode amplitude parameter $A\sim 10^{-3}(\nu_{\rm spin}/10^3{\rm Hz})^{5/2}$, where $\nu_{\rm spin}$
denotes the spin frequency of the star.
If this amplitude estimation is correct,
the amplitude parameter $A$ for neutron stars spinning at $\nu_{\rm spin}\sim 500{\rm Hz}$ becomes
of order of $10^{-4}$, which is by one order of magnitude smaller than the value $A=10^{-3}$ used in this paper 
for presentation and may
suggest that the detectability of the periodic modulation due to the $r$-modes is not very high even for
small values of $\theta_s$.
In this paper we have been discussing the possibility of using X-ray light curves from a hot spot as a probe into the core $r$-modes
of neutron stars, but
it is also conceivable that periodicities due to the $r$-modes, which may shake the surface magnetic field, are contained 
in the pulses, possibly as drifting sub- or micro-pulses, from
millisecond radio pulsars.







\end{document}